\documentclass[11pt]{article}

\usepackage{amscd}
\usepackage{amsmath}

\begin{document}

\newcommand{\N}{N\raise.7ex\hbox{\underline{$\circ $}}$\;$}

\thispagestyle{empty}

\begin{center}
{\bf

BELARUS NATIONAL ACADEMY OF SCIENCES

B.I. STEPANOV's  INSTITUTE OF PHYSICS

BELARUSSIAN STATE UNIVERSITY

 }

\end{center}

\vspace{20mm}

\begin{center}

{\bf  A. A. Bogush,   V.M.  Red'kov\footnote{redkov@dragon.bas-net.by},  G.G.  Krylov\footnote{krylov@bsu.by}}

\end{center}

\vspace{5mm}

\begin{center}

{\bf SCHR\"{O}DINGER PARTICLE IN MAGNETIC AND ELECTRIC FIELDS \\ IN
LOBACHEVSKY AND RIEMANN SPACES
}

\end{center}

\thispagestyle{empty}

 \begin{quotation}

Schr\"{o}dinger equation in Lobachevsky and Riemann  4-spaces has
been solved in the presence of external magnetic field that is an analog
of a uniform magnetic field in the flat space. Generalized Landau
levels have been found, modified by the presence of the space
curvature. In Lobachevsky
4-model the energy spectrum contains discrete and continuous
parts,
  the number of bound states is finite; in Riemann 4-model  all energy spectrum is discrete.
 Generalized Landau levels are   determined by three
parameters, the magnitude of the magnetic field  $B$, the
curvature radius  $\rho$ and the magnetic quantum number   $m$.
It has been shown that in presence of an additional external electric field
the energy spectrum in the Riemann model can be also
  obtained analytically.

\end{quotation}

\vspace{10mm}

{\bf Pacs:} 1130, 0230, 0365

{\bf keywords:} magnetic field, Landau system,  Schr\"{o}dinger equation,
hyperbolical 3-space,

 spherical 3-space, energy levels

\vspace{20mm}
 Published:

\vspace{5mm}

Bogush A.A., Red'kov V.M., Krylov G.G..
Schr\"{o}dinger particle in magnetic and electric fields in
Lobachevsky and Riemann spaces.  Nonlinear Phenomena in Complex Systems. 2008.
Vol. 11. no 4, P.  403 -- 416.

\newpage

\section{Introduction. Spaces of a constant curvature
and uniform magnetic field}


Quantum-mechanical Kepler problems in spaces of constant positive and  negative
curvature were formulated and solved many years ago by Schr\"{o}dinger, Infeld,
 Schild, and Stefenson
\cite{Schrodinger,Infeld,Infeld-Schild,Stevenson}.
Dynamical symmetry for quantum Kepler problems in spaces of constant curvature was
investigated in \cite{Higgs,Leemon,Kurochkin-Otchik,Bogush-Kurochkin-Otchik}.
These models were used to describe  bound state in nuclear physics
 \cite{Izmestiev,Murzov,Murzov'}
 and nano-physics \cite{Kurochkin,Gritzev}.

In the present paper the non-relativistic quantum mechanical problems
for a  particle in a magnetic field on the background of  hyperbolical Lobachebsky and spherical Riemann
spaces  are  solved.

We start with Olevsky  paper  \cite{Olevsky}   where   all 3-orthogonal coordinate systems $(x^{1},x^{2},x^{3})$
in 3-space -f constant curvature,  negative and positive,
permitting the full separation of variables
 in the  extended Helmholtz equation
 $(\Delta_{2} + \lambda ) f(x) = 0 $ have been found.
An  idea consists in
searching for an analog of cylindrical coordinate of Euclid space
$E_{3}$, in which one could easily define the concept of a
uniform magnetic field in spaces $H_{3}$ and $S_{3}$. Such  uniform  magnetic fields
should be solutions of  the corresponding Maxwell equations.
Then we will try to examine the Landau problem
\cite{Landau,Landau-Lifshitz-3}
 of a Schr\"{o}dinger's particle in the external
magnetic field on the  background of curved space models, $H_{3}$ and $S_{3}$.
Also, in  present time we can note growing interest to Landau problem
in the context of development in quantum mechanics in noncommutative space
  \cite{Gamboa-1,Gamboa-2,Gamboa-3}.
In the literature, the Landau problem was examined for Lobachevsky and spherical planes
 \cite{Klauder-Onofri,Klauder-Onofri-et,Dunne,Negro et,Drukker};
 present paper agrees with the  known results and adds to them.

In \cite{Olevsky} in the list of coordinate systems, one can see the system
number  $XI$ defined in space  $S_{3}$ by the relations
\begin{eqnarray}
dS^{2} = c^{2} dt^{2} -  [  \cos^{2} z ( d r^{2} +
\sin^{2} r d \phi^{2} ) - dz^{2} ]\; ,
\nonumber
\\
z \in [ - \pi/2 , + \pi /2 ], \; r \in [0, \pi ] , \; \phi
\in [0, 2 \pi ] \; ,
\nonumber
\\
u_{1} = \cos z \; \sin r \cos \phi \; , \;  u_{2} = \cos z \;
\sin r \sin \phi \; ,
\nonumber
\\
u_{3} = \sin z \; , \;\;   u_{0} = \cos z \cos r \; ,
\;\;
 u_{1}^{2} +  u_{2}^{2} + u_{3}^{2} + u_{0}^{2}  = 1 \; ;
\nonumber
\label{1.1b}
\end{eqnarray}

and in space $H_{3}$ as
\begin{eqnarray}
dS^{2} = c^{2} dt^{2} -  [ \cosh^{2} z ( d
r^{2} + \sinh^{2} r  d \phi^{2} ) - dz^{2} ] ,
\nonumber
\\
z \in ( - \infty , + \infty ), \ r \in [0, +\infty ] , \
\phi \in [0, 2 \pi ] ,
\nonumber
\\
u_{1} = \cosh  z  \sinh r \cos \phi  , \
u_{2} = \cosh  z  \sinh r \sin \phi ,
\nonumber
\\
u_{3} = \sinh z  , \  u_{0} = \cosh  z
\cosh  r  ,
\;\;
u_{1}^{2} +  u_{2}^{2} + u_{3}^{2} - u_{0}^{2}  = -1 \; .
\nonumber
\label{1.1b}
\end{eqnarray}

All coordinate  $u_{a}$ are dimensionless,  what is achieved  by  dividing on the curvature radius $\rho$.
When $\rho \rightarrow \infty$,  these coordinates reduce to usual cylindrical ones  in the space $E_{3}$
\begin{eqnarray}
dS^{2} = c^{2} dt^{2} - (\;  d r^{2} +  r^{2} \; d \phi^{2}  +
dz^{2}\; )\;  .
\nonumber
\end{eqnarray}

Let us use the known form of the vector potential of the  uniform constant magnetic field [2]
in the flat space:
\begin{eqnarray}
{\bf A} = {1 \over 2} \; {\bf B} \times {\bf r} \; , \qquad {\bf B} =
(0,0,B) \; ,
\qquad A^{a} =  {B r \over 2} (0,
-\sin \phi, \cos \phi, 0) \;,
\label{1.2a}
\end{eqnarray}
\noindent after translating to cylindrical coordinates
\begin{eqnarray}
A_{t} = 0, \; A_{r} =  0 \; , \; A_{z} =
0 \; , \;
A_{\phi} = -{Br^{2} \over 2} \; .
\label{1.2b}
\end{eqnarray}
\noindent Correspondingly we have a single non-vanishing component of the strength tensor
which obeys the Maxwell equations:
\begin{eqnarray}
F_{\phi r} =
\partial_{\phi}A_{r} - \partial_{r} A_{\phi} = B r \; ,
\qquad
 {1 \over r }
{\partial \over \partial r } r  F^{\phi  r} = {1 \over r }
{\partial \over \partial r } r \; (- {1 \over r^{2}} )\;Br  \equiv
0 \; .
\label{1.3a}
\end{eqnarray}

Such description of a uniform magnetic field in cylindrical coordinates
 may be extended quite easily to curved  space  models.
Let us start with the following potential in the spherical space $S_{3}$
\begin{eqnarray}
A_{\phi} = 2B\; (\cos r -1 )\;  , \;
F_{\phi r} =
B   \sin r \; ,
\label{1.4a}
\end{eqnarray}
\noindent
Analogously,  in Lobachevsky model we start with $H_{3}$
\begin{eqnarray}
A_{\phi} =  -2B\; ( \cosh\; r
-1 )\; , \qquad
F_{\phi r} = B  \; \sinh\; r \; .
\label{1.5a}
\end{eqnarray}

\noindent As one can see both potentials are solutions of the respective  Maxwell equations,
and in the vanishing curvature limit they are reduced to (\ref{1.3a}).

\section{  Separation of variables in the Schr\"{o}dinger equation
}

Non-relativistic Schr\"{o}dinger equation in a space-time with
the metrics
\begin{eqnarray}
dS^{2} = (dx^{0})^{2} + g_{kl}(x) dx^{k}dx^{l} \; , \; g =
\mbox{det} \; (g_{kl}(x))
\nonumber
\end{eqnarray}
\noindent has the form
\begin{eqnarray}
( \;  i  \hbar  \;  \partial  _{t }   + e  \; A_{0} \; ) \;   \Psi
= H \Psi \; ,  \qquad
\nonumber
\\[2mm]
 H={-1 \over 2M}\;  [\;   ({i \hbar  \over \sqrt{-g} } \;
\partial_{k} \sqrt{-g}  + {e \over c}   A_{k} ) g^{kl}
(i \hbar  \partial  _{l } +  {e\over c } A_{l})  \;  ]\;     \Psi .
\nonumber
\label{2.1}
\end{eqnarray}
In the space  $S_{3}$,  parameterized by cylindric coordinates and in the presence of the external uniform
magnetic field, the Hamiltonian   $H$ takes the form
\begin{eqnarray}
H = { \hbar^{2} \over 2MR^{2} }  \left[ - {  1\over \cos^{2}z \sin
r}   {\partial \over  \partial r }  \sin r
  {\partial \over  \partial r }
  +
{1 \over \cos^{2}z \sin^{2} r   }    \left(i {\partial \over
\partial \phi  } + {e\over \hbar c} A_{\phi} \right) ^{2} -  {1 \over
\cos^{2} z } {\partial \over  \partial z}  \cos^{2} z { \partial
\over \partial z}  \right]
\nonumber
\label{2.2}
\end{eqnarray}
\noindent
and the variables may be separated by the substitution:
\begin{eqnarray}
\Psi = e^{-iEt / \hbar } e^{im \phi } \; Z(z)\; R(r) \; , \qquad
\epsilon = E / (\hbar^{2} / M\rho^{2} ) \; ,
\label{2.3}
\\
{1 \over Z}  \; \left( 2\epsilon   \cos ^{2} z \; Z   +    {d
\over  d z}  \cos^{2} z { d Z \over d z}    \right)
\nonumber
\\
= {1 \over R}  \;  \left( - {  1\over  \sin r}   { d \over  d r }
\sin r
  {d R \over  d r }   +
 {1 \over \sin^{2} r   }    (-m  + {e\over \hbar c} A_{\phi} ) ^{2}  R
 \right) \; .
\nonumber
\end{eqnarray}
\noindent Introducing the separation constant   $\lambda$, we arrive at
\begin{eqnarray}
   {d \over  d z}  \cos^{2} z { d Z \over d z}  + 2\epsilon   \cos ^{2} z
   Z     =
\lambda  Z  ,
\label{2.4a}
\\
- {  1\over  \sin r}   { d \over  d r }  \sin r
  {d R \over  d r }   +
 {1 \over \sin^{2} r   }
(m  +   {eB \over \hbar c} \; 2\rho^{2} \sin^{2} {r \over 2}  ) ^{2}  R
   = \lambda  R  .
 \label{2.4b}
\end{eqnarray}

\noindent
In the same manner, in the Lobachevsky space $H_{3}$
we arrive at
\begin{eqnarray}
   {d \over  d z}  \cosh^{2} z { d Z \over d z}  + 2\epsilon
    \cosh^{2} z  Z     =
\lambda Z  ,
\label{2.5a}
\\
- {  1\over  {\sinh}  r}   { d \over  d r }  {\sinh}  r
  {d R \over  d r }   +
 {1 \over {\sinh}^{2} r   }
 \left(m  +   {eB \over \hbar c} 2\rho^{2} {\sinh}^{2} {r \over 2}
 \right) ^{2}  R
  = \lambda R  .
 \label{2.5b}
\end{eqnarray}
\noindent
In the limit of  flat space these equations become more simple:
\begin{eqnarray}
   { d^{2} Z \over d z^{2}}  +  {2EM\over \hbar^{2}}  Z    =
\Lambda  Z, \
 Z = e^{\pm i P z/\hbar}  ,
 \Lambda = {2M \over \hbar^{2}}
 \left(E - {P^{2} \over 2M} \right) ,
 \nonumber
 \\
 - {  1\over   r}   { d \over  d r }   r
  {d R \over  d r }
   +
 {1 \over  r ^{2}  }    \left(m  +   {eB \over \hbar c} {r^{2}\over 2}
 \right) ^{2} R
   = {2M \over \hbar^{2}} \left(E - {P^{2} \over 2M} \right)  R
 \label{2.6b}
\end{eqnarray}
\noindent which coincide with the  known one
 \cite{Landau-Lifshitz-3}:
\begin{eqnarray}
{\hbar^{2} \over 2M} \left(  {d^{2} R \over d r^{2}} -{1 \over r}
{d R \over d r} -{m^{2} \over r^{2}}  R\; \right)
 + \left[  E
-{P^{2} \over 2M}  - {\omega^{2} \over 8 } r^{2}  - {\hbar \omega
m \over 2}  \right]  R = 0
\nonumber
\end{eqnarray}
\noindent with  $\omega = eB / Mc$.

\vspace{0.4cm}\section{ Radial solutions in space    $H_{3}$}

Below the notion ${eB \over \hbar c} \rho^{2}  \Longrightarrow  B $ will be  used, then
 eq.   (\ref{2.5b}) reads
\begin{eqnarray}
{d^{2} \over dr^{2}} R + {1 \over {\tanh} r } {d R \over d r }
  -
 {1 \over {\sinh}^{2} r   }    [\;  m  +   B  (\cosh r -1)] ^{2}
  R
    + \lambda   R = 0 .
\label{3.1}
\end{eqnarray}
\noindent
In new variables
\begin{eqnarray}
\cosh r -1 = -2z , \
 z = - {\sinh}^{2} {r \over 2} \in ( -\infty , 0  ]  ,
\nonumber
\end{eqnarray}
\noindent for the radial equation we get
\begin{eqnarray}
\left [ z (1-z) {d^{2} \over dz^{2}} +  (1-2z)  {d \over dz}
-
 {1 \over 4} ( {m^{2}\over  z} -4B^{2} + {(m-2B)^{2}\over 1-z}  )
-\lambda  \right ] R = 0  .
\label{3.4}
\end{eqnarray}

Let us specify the behavior of the function  $R$ near to  the points $0$  and $1$ (take note that the value
 1 does not belong to the physical domain of the coordinate $r$):
\begin{eqnarray}
\mbox{if } z  \longrightarrow   0 , \ \mbox{then }
{ R(z)  \sim  z^{a} } , \
a = \pm { m  \over 2} ;
\nonumber
\\
\mbox{if } z  \longrightarrow  1  , \ \mbox{then }
R (z) \sim (1-z)^{b} , \            b =  \pm {m-2B  \over 2}   .
\nonumber
\label{3.5b}
\end{eqnarray}
\noindent
Making  substitution
$
R = z^{a} (1-z)^{b}  F  $,
 from eq. (\ref{3.4}) we get
\begin{eqnarray}
+ z (1-z) \; F'' + [\;  a (1-z)  - b z   + a  (1-z)   - b z  +
(1-2z)  ] \; F'  +
\nonumber
\\
+ {1 \over z}\;  [ \; a (a-1)  + a  -  {m^{2} \over 4 } \;] \;  F
\;\; + \;\;
 {1 \over 1-z }\; [\;   b (b-1)  + b  - {(m-2B)^{2} \over 4 }  \;]\;  F -
\nonumber
\\
- [  a(a+1)  + 2a b         +   b (b+1)
  - B^{2}     + \lambda  ] \;  F = 0 \; ;
\label{3.6a}
\end{eqnarray}
\noindent Imposing conditions
$
a = \pm  m   /  2    , \; b = \pm (m-2B )/ 2
$ we obtain
\begin{eqnarray}
R = z^{a}  (1-z)^{b} F ,
\nonumber
\\
 z (1-z) F'' + [(2a+1) -2 (a+b+1)z ]  F'
\nonumber
\\
- [  a(a+1)  + 2a b         +   b (b+1)
  - B^{2}     + \lambda  ]   F = 0
\nonumber
\label{3.6c}
\end{eqnarray}
\noindent which is the  equation of hypergeometric  type:
$z(1-z)  F + [ \gamma - (\alpha + \beta +1) z ]  F' - \alpha
\beta  F = 0  .
$
Thus, the radial function is constructed in hypergeometric functions as follows:
(to obtain the bound states we must take the parameters $a$ and
 $b$ positive  and negative respectively)
\begin{eqnarray}
z = - \sinh^{2}{r \over 2}  ,   z \in  ( - \infty, +
0 ] ,  \nonumber \\
R = (- \sinh  {r \over 2})^{\mid  m  \mid }   (\cosh
{ r \over 2} )^{-\mid m -2B \mid}
{_2}F_{1}(\alpha , \beta; \gamma;
- {\sinh}^{2}{r \over 2} )
\label{3.7}
\end{eqnarray}
\noindent where  $(\alpha , \beta)$ are defined by
\begin{eqnarray}
 \alpha + \beta  = 2a + 2b +1  , \qquad  \alpha \beta  =  (a+b)(a+b+1) - B^{2}     + \lambda  \
\nonumber
\end{eqnarray}

\noindent that is
\begin{eqnarray}
a = + {\mid m \mid
\over 2 }, \ b =  - { \mid m-2B  \mid  \over 2}  ,
\nonumber
\\
\alpha = a + b +{1\over 2} - \sqrt{B^{2}  +{1 \over 4} - \lambda} ,
\nonumber \\
 \beta = a + b +{1\over 2} + \sqrt{B^{2}  +{1 \over 4} - \lambda},
\nonumber
\\
\gamma = 2a + 1 =  + \mid m \mid +1  .
\label{3.8}
\end{eqnarray}
The first possibility to obtain polynomial is
\begin{eqnarray}
\alpha = a + b +{1\over 2} - \sqrt{B^{2}  +{1 \over 4} - \lambda
} = -n  = 0, -1,\ldots
\nonumber
\end{eqnarray}
\noindent from whence it follows the quantization rule
\begin{eqnarray}
a+b + {1\over 2} + n  \geq 0     ,
\qquad
 \lambda  = {1 \over 4} +  B^{2}   - \left(a+b  + n +{1\over 2}\right)^{2}
\label{quant1}
\end{eqnarray}
\noindent or
\begin{eqnarray}
{\mid m \mid \over 2} - {\mid m -2B \mid \over 2} + n +{1 \over 2} \geq 0  ,
\nonumber
\\
\lambda = {1 \over 4 } + B^{2} - ( {\mid m \mid \over 2} - {\mid m
-2B \mid \over 2} + n +{1 \over 2})^{2} ,
\label{3.9}
\end{eqnarray}
\begin{eqnarray}%
&R = (- \sinh {r \over 2})^{\mid  m  \mid }  (\cosh{ r \over 2} )^{-\mid m -2B \mid}
 {_2}F_{1}(-n , \mid m \mid - \mid m - 2B \mid +1 +n ; \mid m
\mid +1,    - {\sinh}^{2}{r \over 2} )
.
\end{eqnarray}
\noindent
The second possibility to obtain the polynomial is
\begin{eqnarray}
&\beta = a + b +{1\over 2} + \sqrt{B^{2}  +{1 \over 4} - \lambda }
= -n
\nonumber
\end{eqnarray}
\noindent that is
\begin{eqnarray}
a+b + {1\over 2} + n \leq 0     ,
\qquad &\lambda  =  {1 \over 4} + B^{2}  - \left(a+b +{1\over 2} + n\right)^{2}
\label{quant2}
\end{eqnarray}
\noindent
or
\begin{eqnarray}
& {\mid m \mid \over 2} - {\mid m -2B \mid \over 2} + n +{1 \over 2} \leq 0  ,
 \nonumber \\
&\lambda = {1 \over 4 } + B^{2} - \left( {\mid m \mid \over 2} - {\mid m
-2B \mid \over 2} + n +{1 \over 2}\right)^{2} ,
\label{3.10}
\end{eqnarray}
\begin{eqnarray}
R = (- \sinh\; {r \over 2})^{\mid  m  \mid }  (\cosh
{ r \over 2} )^{-\mid m -2B \mid}
{_2}F_{1}(\mid m \mid - \mid m - 2B \mid +1 +n ,  -n; \mid m
\mid +1  ;   - \sinh^{2}{r \over 2} )  .
\end{eqnarray}

Now, let we assume that $B >0$. There exists five possibilities:
\begin{eqnarray}
     m= 0  ,
   m < 0   , \
    0 < m < 2B  ,\
    m > 2B,    m = 2B  .
\nonumber
\end{eqnarray}
For the case $m=0$ we have
$ \ a=0 , \ b = - B .
$
\noindent Applying  the quantization condition   (\ref{3.9}) we get a very special situation: namely,
one separate bound state may exist or not depending on the value of a magnetic field:
\begin{eqnarray}
0 < B-n \leq  {1 \over 2}
, \qquad
 \lambda -{1\over 4}  = 2B(n+ 1 /2 ) - (n+ 1 / 2)^{2}  ,
\nonumber
\\
R = (\cosh { r \over 2} )^{-2B } \; {_2}F_{1}(-n , -  2B  +1 +n ;
+1  ;   - \sinh^{2}{r \over 2} )  ,
\nonumber
\\
 R _{r \rightarrow +\infty} \sim
(\cosh { r \over 2} )^{-2B}  (\sinh^{2}{r \over 2}
)^{n}
\sim e^{- (B  - n )r  }  .
\label{3.12}
\end{eqnarray}
Applying the quantization rule (\ref{3.10}) we get
a  finite series of bound states:
\begin{eqnarray}
n+ {1 \over 2 }  \leq B  ,
\ n= 0,1,2, \ldots , N_{B}  ,
\qquad
 \lambda  -1/4 = 2B(n+1/2) - (N+1/2)^{2}  ,
\nonumber
\\ \nonumber
R =  (\cosh { r \over 2} )^{-2B}
{_2}F_{1}(  -  2B  +1 +n ,
-n; \mid m  \mid +1  ;  - \sinh^{2}{r \over 2} )  ,
\nonumber
\\
R _{r \rightarrow +\infty} \sim (\cosh  { r \over 2} )^{-2B}
(\sinh^{2}{r \over 2} )^{n} \sim e^{- (B  - n )r  }  .
\label{3.13}
\end{eqnarray}

Now let us consider the case
\begin{eqnarray}
m < 0\;  ,  \;\;\; a = - {m \over 2} > 0
\; , \;\;\;  b = {m\over 2} - B  < 0 \; .
\nonumber
\end{eqnarray}
\noindent Applying  the quantization condition   (\ref{3.9}) we again get a  special situation:
one separate bound state may exist or not depending on the value of magnetic field:
\begin{eqnarray}
 0 < B -n  \leq {1
\over 2} \; ,
\qquad \lambda - {1 \over 4 } =  2B(n+1/2) - (n+1/2)^{2}\; ,
\nonumber
\\
R = (- \sinh\; {r \over 2})^{-   m   }  (\cosh\; { r
\over 2} )^{ m -2B }
 {_2}F_{1}(-n , - 2B  +1 +n ; -  m +1  ; \;  - \sinh^{2}{r
\over 2} ) \; ,
\nonumber
\\
R_{r \rightarrow +\infty} \sim e^{-(B-n) r } \; .
\label{3.15}
\end{eqnarray}
Applying the quantization rule (\ref{3.10}) we get
a finite series of bound states:
\begin{eqnarray}
n +{1 \over 2} \leq B
, \ n = 0,1, 2, \ldots,  N_{B}  ,
\nonumber
\\
\lambda  -1/4 = 2B(n+1/2) - (n+1/2)^{2} ,
\nonumber
\\
R = (- \sinh\; {r \over 2})^{-  m  }  (\cosh\; { r
\over 2} )^{ m -2B } {_2}F_{1}( -2B  +1 +n ,  -n;  - m  +1  ;   - \sinh^{2}{r \over 2}
)  ,
\nonumber
\\
R_{r \rightarrow +\infty} \sim e^{-(B-n) r }  .
\label{3.16}
\end{eqnarray}

Now let us consider the case
\begin{eqnarray}
 0 < m < 2B  , \ a =  {m \over 2} >
0   , \ b = {m \over 2}  -B  < 0  .
\nonumber
\end{eqnarray}
\noindent Applying  the quantization condition   (\ref{3.9}) we get a very special situation:
one separate bound state may exist or not depending on the value of magnetic field
and magnetic quantum number:
\begin{eqnarray}
0 \leq B - m - n
\leq {1 \over 2} ,
\qquad
\lambda -1/4 = 2B (m+ n +1/2) -(m + n +1/2)^{2}  ,
\nonumber
\\
R = (- \sinh {r \over 2})^{  m  } \; (\cosh\; { r
\over 2} )^{ m -2B }  {_2}F_{1}(-n , 2  m  - 2B  +1 +n ;  m   +1  ;   - \sinh^{2}{r \over
2} ) ,
\nonumber
\\
R_{r \rightarrow + \infty} \sim e^{-(B- m  - n )r }  .
\label{3.18}
\end{eqnarray}
Applying the quantization rule (\ref{3.10}) we get
a finite series of bound states:
\begin{eqnarray}
m   + {1\over 2} + n  \leq B  \;  , \qquad n = 0, 1, ..., N_{B,m}
\; ,
\nonumber
\\
\lambda -1/4 = 2B(m+n+1/2) - (m+n+1/2)^{2}  \;,
\nonumber
\\
R = (- \sinh\; {r \over 2})^{  m   } (\cosh\; { r
\over 2} )^{ m -2B }
 {_2}F{_1} ( 2  m  - 2B +1 +n ,  -n;   m   +1  ;
 - \sinh^{2}{r \over 2} ) \; ,
 \nonumber
 \\
 R_{r \rightarrow + \infty} \sim e^{-(B- m  - n )r } \; .
\label{3.19}
\end{eqnarray}

For the case
\begin{eqnarray}
m > 2B 
 \qquad a = {m \over 2} \; , \qquad b = -{m -2B \over 2}  < 0 \; ,
\nonumber
\end{eqnarray}
\noindent with the rule  (\ref{3.9}) we get
\begin{eqnarray}
B  + {1\over 2} +n  \geq 0     , \qquad
 \lambda - {1 \over 4 } = -2B (n+ {1 \over 2}) - (n +{1 \over 2})^{2}  ,
\nonumber
\\
R = (- \sinh\; {r \over 2})^{  m  } (\cosh { r
\over 2} )^{-  m +2B }   {_2}F_{1}(-n , 2B  +1 +n ;  m  +1  ;   - \sinh^{2}{r
\over 2} ) ,
\nonumber
\\
R_{r \rightarrow + \infty} \sim e^{(B+n)r }   \rightarrow   \infty  ;
\label{3.21}
\end{eqnarray}
\noindent this is not the   case of bound states.
With the rule  (\ref{3.10}) we arrive at
\begin{eqnarray}
B + {1 \over 2} +
n  \leq 0 \;\;
\end{eqnarray}
\noindent  that can not be fulfilled together with the
quantization conditions whereas $n\ge 0$ .

And finally for the  last variant:
\begin{eqnarray}
m = 2B  > 0 \;  , \qquad a = m = 2B \; ,
\qquad b = 0 \; ,
\nonumber
\end{eqnarray}

\noindent  we get
\begin{eqnarray} \nonumber
B  + n +{1 \over 2} \geq 0 \; , \qquad
 \lambda - {1 \over 4 }  = 2B(n+{1 \over 2}) - ( n +{1 \over
2})^{2}\;,
\nonumber
\end{eqnarray}
\begin{eqnarray} \nonumber
R = (- \sinh {r \over 2})^{ 2B } \ 
{_2}F_{1}(-n , \mid m \mid
-\mid
m - 2B \mid +1 +n ; \mid m  \mid +1  ;   - \sinh^{2}{r \over
2} )  , %
\nonumber \\
 R_{r \rightarrow +\infty } \sim e^{(B+n)r}  \rightarrow \infty
\nonumber\
\end{eqnarray}
\noindent which is not a bound state. Quantization according to  (\ref{3.10})
cannot provide us with bound states because
of relationship
$
B  + n + 1 / 2 > 0 $ .

Let us collect the results together:

\vspace{2mm}
\noindent
\underline{separate  states}:
that may be  presented by the single formula:
\begin{eqnarray}
 m < 2B  , \ n < B \leq n+1/2 ,\  n=0,1,2,\ldots ,
\nonumber
\\
\lambda -{1\over 4} = 2B( {m + \mid m \mid \over 2 } +n+1/2)
 - \left(  { m +
\mid m \mid \over 2} + n+{1\over 2}\right)^{2}  ;
\label{3.25b}
\end{eqnarray}

\noindent
\underline{series of states }:
given by a  formula
\begin{eqnarray}
m < 2B  , \ {m +\mid m \mid \over 2} + n
+1/2  \leq B   ,
\nonumber
\\
\lambda -1/4 = 2B\left( {m + \mid m \mid \over 2 } + n + 1/2\right)
- \left(  {
m + \mid m \mid \over 2} + n + {1\over 2}
\right)^{2},\  n =0,1,\ldots , N_{B}  .
\nonumber
\\
\label{3.26b}
\end{eqnarray}
\noindent
In usual unites, the  last relation reads
\begin{eqnarray}
\lambda - {1\over 4} = \rho^{2}  \lambda_{0} - {1 \over 4}  ,
\ \lim_{\rho \rightarrow \infty} \;\lambda_{0} = {2M \over
\hbar^{2}}  ( E -{P^{2} \over 2M}) ,
\qquad
m < 2B  , \ m+ n +1/2  \leq  {eB  \over \hbar c}
\rho^{2}  ,
\nonumber
\\
\rho^{2}  \lambda_{0} - {1 \over 4} = 2 {eB  \over \hbar c}
\rho^{2} ( {m + \mid m \mid \over 2 } + n + 1/2)
 - (  { m + \mid
m \mid \over 2} + n + 1/2)^{2},\  n =0,1,\ldots, N_{B} .
\nonumber
\\
\label{3.27}
\end{eqnarray}
\noindent In the limit of a flat space, from (\ref{3.27}) the known result follows:
\begin{eqnarray}
 E -{P^{2} \over 2M} =
 \; {eB \hbar \over M c} \; ( {m + \mid m \mid \over 2 } + n + 1/2) \; .
\nonumber
\end{eqnarray}

For the case of the  positive  orientation of magnetic field $
B < 0\; $,  we should take into account  the symmetry provided by initial differential equation:
$
m \rightarrow  m' = - m\; , \;
B \rightarrow B' = - B \; .
$

\section{ Differential equation for  $Z(z)$  in  $H_{3}$}

Now, let us examine the differential equation for
$Z(z)$  in space $H_{3}$:
\begin{eqnarray}
{d^{2} Z\over dz^{2}}  +  2\;  { \sinh z \over \cosh z
} {d Z \over dz}    + 2\epsilon  Z - {\lambda  \over
\cosh^{2} z}  Z = 0  .
\label{4.1}
\end{eqnarray}
\noindent Changing the  variable as $\sinh\; z = i (2x-1) $, we get
\begin{eqnarray} \nonumber
x(1-x)  {d^{2} Z \over d x^{2}} - 3  (2 x-1)  {1 \over 2} \; {d
Z\over d x}   - \left( 2\epsilon -{ \lambda \over  4x(1-x)} \right)\; Z = 0
.
\label{4.5}
\end{eqnarray}
Taking the function in the form
$
Z = x^{a} (1-x)^{b} F \; ,
$ we obtain
\begin{eqnarray}
+ x (1-x) F'' +
 [ 2a  (1-x)   - 2b x      -3x  +  {3 /
 2} ]   F'
+ \left[  a (a-1)+  {3 a\over 2}   + {\lambda \over 4}  \right]
{1\over x} F
\nonumber
\\
+ \left[   b(b-1) +3b  -  {3 b\over 2} + {\lambda
\over 4}  \right] {1 \over 1-x}  F
+ [ -a(a-1)  - a b   - ab     - b(b-1)    - 3 a    - 3b
- 2\epsilon ]   F    = 0  .
\label{4.7a}
\end{eqnarray}
\noindent Imposing additional restrictions
\begin{eqnarray}
a^{2} + {a\over 2} + {\lambda \over 4}= 0 , \ a = {-1 \pm
\sqrt{1-4\lambda} \over 4} , \qquad b^{2} + {b\over 2} + {\lambda \over 4}= 0 ,\
b = {-1 \pm \sqrt{1-4\lambda} \over 4} ,
\label{4.7b}
\end{eqnarray}
\noindent we arrive at  more simple  equation
\begin{eqnarray}
x (1-x)  F'' +
 [   (2a  +3/2) -x (2a +2b +3)  ]   F'
 - [ (a+b+1)^{2}     + 2 \epsilon  -1 ]   F    = 0
\label{4.8a}
\end{eqnarray}
\noindent which is of hypergeometric type
\begin{eqnarray}
x(1-x) F + [ C - (A + B +1) x ] F' - AB   F = 0
\nonumber
\end{eqnarray}
\noindent with parameters defined by
\begin{eqnarray}
&A + B = 2(a+b+1)  , \qquad
AB = (a+b+1)^{2} +2 \epsilon -1  .
\nonumber
\end{eqnarray}
\noindent That is
\begin{eqnarray}
A = a +b + 1 + \sqrt{1-2\epsilon} ,\qquad
B = a +b + 1 - \sqrt{1-2\epsilon} ,
\;
C = 2a + 3/2  ;
\nonumber
\\
Z = x^{a} (1-x)^{b} {_2}F_{1}(A, B; C; x)   , \
 x = {1 -\imath \sinh z \over 2},
 \;\; (1- x) = {1 + \imath  \sinh z \over 2} ,
\nonumber
\\
 a = {-1 \pm \sqrt{1-4\lambda} \over 4} , \
 b = {-1 \pm \sqrt{1-4\lambda} \over 4}  .
 \label{4.8c}
\end{eqnarray}
To find behavior of these solutions at infinity,
one can use the Kummer's relations
\cite{Bateman}
\begin{eqnarray}
U_{1} ={_2}F_{1}(A, B; C; x)  , \qquad
\nonumber
\\
U_{3} = (-x)^{-A }  {_2}F_{1}(A, A - C +1; A - B +1; x^{-1} )  ,
\nonumber
\\
U_{4} = (-x)^{-B} {_2}F_{1}(B -C +1,  B; B - A +1;  x^{-1})  ,
\nonumber
\\
U_{1} =  { \Gamma (C) \Gamma (B - A) \over \Gamma(C - A) \Gamma
(B) }  U_{3} + {\Gamma (C) \Gamma (A - B) \over \Gamma(C - B)
\Gamma (A) }  U_{4} .
\label{4.9a}
\end{eqnarray}
Then
\begin{eqnarray}
Z = x^{a} (1-x)^{b} U_{1}
=  \hspace{30mm}
\nonumber \\
{ \Gamma (C) \Gamma (B - A) \over \Gamma(C - A) \Gamma (B) }
x^{a} (1-x)^{b}\;\; (-x)^{-(a+b+1 + i \sqrt{2\epsilon -1}) }{_2}F_{1}(A,
A - C +1; A - B +1;x^{-1} )
\nonumber
\\
+ {\Gamma (C) \Gamma (A - B) \over \Gamma(C - B) \Gamma (A) }
x^{a} (1-x)^{b}\;\; (-x)^{-(a+b+1 - i \sqrt{2\epsilon -1})}
{_2}F_{1}(B - C +1,  B; B - A +1;  x^{-1})  . \nonumber
\\
\label{4.9b}
\end{eqnarray}
\noindent
Therefore, asymptotically all three solutions   $U_{1},U_{3},U_{4}$ vanish when
 $x \rightarrow \infty$:
\begin{eqnarray}
Z = x^{a} (1-x)^{b} U_{1}
\sim
\nonumber
\\
\sim
{ \Gamma (C) \Gamma (B - A) \over \Gamma(C - A) \Gamma (B)
}  x^{a} (1-x)^{b} (-x)^{-a-b-1 -i\sqrt{2\epsilon -1 } }  +
 {\Gamma (C) \Gamma (A - B) \over \Gamma(C - B) \Gamma (A) }
x^{a} (1-x)^{b}
\nonumber
\\
\times (-x)^{-a-b-1 + i \sqrt{2\epsilon-1}}
=
\nonumber
\\
 (-1)^{a} \left [  { \Gamma (C) \Gamma (B - A) \over
\Gamma(C - A) \Gamma (B) }  (-x)^{-1 -i\sqrt{2\epsilon-1} }  +
{\Gamma (C) \Gamma (A - B) \over \Gamma(C - B) \Gamma (A) }
(-x)^{-1 +\imath \sqrt{2\epsilon-1}}  \right  ]
 \rightarrow
 0  .
\nonumber
\\
\label{4.9c}
\end{eqnarray}
\vspace{15mm}
Let  $a = b$,  then
\begin{eqnarray}
a=b = {-1 \pm 2i \sqrt{\lambda-1/4 } \over 4}   , \ a+b  =
{-1 \pm 2i \sqrt{\lambda -1/4} \over 2}   ,
\nonumber
\\
Z = x^{a} (1-x)^{a}\; {_2}F_{1}(A, B; C; x)
=  ({\cosh^{2} z \over 4}
)^{a}  \; {_2}F_{1}(A, B; C; x) ,
\nonumber
\\
A =  {1 \over 2}  +  \imath  (  \sqrt{2\epsilon-1} \pm
\sqrt{\lambda-{1\over 4}}  )  ,\;
B =  {1 \over 2}   -  \imath (  \sqrt{2\epsilon-1}  \mp
\sqrt{\lambda-{1\over 4}}  ) , \nonumber
\\
  C =  1 \pm \imath \sqrt{\lambda -1/4} .
\label{4.11}
\end{eqnarray}
One may note that from  (\ref{4.11}) it follows
\begin{eqnarray}
(A-{1 \over 2} )  (B-{1 \over 2} ) =
(2\epsilon-1)- (\lambda-{1/
4}) \; .
\nonumber
\end{eqnarray}
\noindent It has sense  to introduce a new parameter
\begin{eqnarray}
(2\epsilon-1) - (\lambda-{1/
4}) = \nu^{2} \; , \qquad \mbox{or} \qquad
 (\lambda-{1/
4}) = (2\epsilon -1) -  \nu^{2}
\label{4.12}
\end{eqnarray}
\noindent which generalizes the known relation in the flat space model
$
\lambda = 2\epsilon  - k^{2}      .
$
It is the matter of simple calculation to determine the operator
with eigenvalues  $\nu^{2} +3/4$:
\begin{eqnarray}
\hat{\nu}^{2} +{3\over 4} = 2 \hat{H} (r,\phi,z)   - \hat{\lambda}
(r) .
\nonumber
\end{eqnarray}
\noindent Indeed, with the notion
\begin{eqnarray}
\hat{\lambda} (r, \phi) = -
{1 \over  \sinh r}   {\partial \over  \partial r }
\sinh  r
  {\partial \over  \partial r }
  +
{1 \over \sinh^{2} r   }    \left[\imath {\partial \over  \partial
\phi  } -
 B (\cosh r -1)   \right] ^{2}  ,
\nonumber
\end{eqnarray}
\begin{eqnarray}
2\hat{H}(r, \phi, z)   = {  1\over \cosh^{2}z }
\hat{\lambda} (r,\phi)
 - {1 \over \cosh^{2} z } {\partial
\over  \partial z}  \cosh^{2} z { \partial \over \partial z}
.
\label{4.14}
\end{eqnarray}

\noindent We  get
\begin{eqnarray}
\hat{\nu}^{2}(r,\phi,z)   +{3\over 4} = \left({  1\over \cosh^{2}z
} - 1\right) \hat{\lambda} (r,\phi) - {1 \over \cosh^{2} z }
{\partial \over  \partial z}  \cosh^{2} z { \partial \over
\partial z}
 .
\label{4.15}
\end{eqnarray}
\noindent and the commutative relations $[ 2 H  , \
 \hat{\lambda} (r) ] =0  $ holds.

\section{  Radial equation in spherical space }

Now let us consider the radial equation arising in spherical space. With the use of
dimensionless parameter $
{eB \over \hbar c}  \rho^{2}  \Longrightarrow   B $
the  equation reads
\begin{eqnarray}
{d^{2} \over dr^{2}} R + {1 \over \tan r } {d R \over d r }
 -
 {1 \over \sin^{2} r   }    [  m  +   B  (1 -  \cos r )] ^{2} R
   + \lambda \   R = 0.
\label{5.1b}
\end{eqnarray}
\noindent
With  variable change given by $1 - \cos r  = 2z $ we have
\begin{eqnarray}
 [ z (1-z) {d^{2} \over dz^{2}} +  (1-2z)  {d \over dz}
-
 {1 \over 4} ( {m^{2}\over  z} -4B^{2} + {(m+2B)^{2}\over 1-z}  )
+\lambda  ] R = 0  .
\label{5.4}
\end{eqnarray}
\noindent
After substitution
$
R = z^{a} (1-z)^{b}  F
$
 we produce
\begin{eqnarray}
 z (1-z) F'' + [ a (1-z)  - b z   + a  (1-z)   - b z  +
(1-2z)  ]  F'
+ {1 \over z} [  a (a-1)  + a  -  {m^{2} /
4 } ]   F
\nonumber
\\
 +
 {1 \over 1-z } [   b (b-1)  + b  - {(m+2B)^{2} /
 }  ] F
- [   a(a+1)  + 2a b         +   b (b+1)
  - B^{2}     - \lambda  ]  F = 0  .
\label{5.5}
\end{eqnarray}

\noindent Imposing the conditions
$a = \pm  \;  m /  2  \;  , \; b = \pm ( m+2B )/  2 $
we arrive at
\begin{eqnarray}
 z (1-z)  F'' + [ (2a+1) -2 (a+b+1)z]   F'
 - [  a (a+1)  + 2a b         +   b (b+1)
  - B^{2}     - \lambda  ]   F = 0 \ \quad
\label{5.6b}
\end{eqnarray}
\noindent which is of hypergeometric type.
 Thus, the problem is solved as
\begin{eqnarray}
z =  \sin^{2}{r \over 2}, \ z \in  [ 0 ,  +
1   ] , \ r \in [  0 , + \pi   ]  ;
\nonumber
\\
R = ( \sin {r \over 2})^{+\mid  m  \mid }
(\cos\; { r \over 2} )^{+ \mid m +2B \mid}
 {_2}F_{1}(\alpha ,
\beta; \gamma ;  - \sin^{2}{r \over 2} )
\label{5.7}
\end{eqnarray}
\noindent where
\begin{eqnarray}
 a = + {\mid m \mid \over 2
}, \ b =  + { \mid m+2B  \mid  \over 2}  , \qquad
\alpha = a + b +{1\over 2} - \sqrt{B^{2}  +{1 \over 4} + \lambda}\; ,
\nonumber
\\
\beta = a + b
 +{1\over 2} + \sqrt{B^{2}  +{1 \over 4}
+ \lambda},\qquad  \gamma =  + \mid m \mid +1 .
\label{5.8}
\end{eqnarray}

The quantization condition is given by
\begin{eqnarray}
\alpha = a + b +{1\over 2} - \sqrt{B^{2}  +{1 \over 4} + \lambda
} = -n  , \qquad
 n  = 0, -1, \ldots
\nonumber
\end{eqnarray}
\noindent from whence it follows relations for possible bound states:
\begin{eqnarray}
\lambda   + {1 \over 4}  = - B^{2} + (a+b +{1\over 2} + n)^{2}
, \qquad \ n  = 0, -1, -2, \ldots
\nonumber
\\
R = ( \sin {r \over 2})^{+\mid  m  \mid }
(\cos { r \over 2} )^{+ \mid m +2B \mid}
 {_2}F_{1}(-n  ,  \mid  m  \mid  +  \mid m +2B \mid +1  + n
; \mid m \mid +1 ;   - \sin^{2}{r \over 2} ).  ,  .
\label{5.9b}
\end{eqnarray}

While examining wave functions in spherical space one must
take into account peculiarities  in parameterization of  the space $S_{3}$  by
the coordinates $(r,\phi,z)$:
\begin{eqnarray}
z \in [ - \pi/2 , + \pi /2 ]  , \ r \in [0, \pi ] ,
\ \phi \in [0, 2 \pi ]  ,
\nonumber
\\
u_{1} = \cos z  \sin r \cos \phi  , \  u_{2} = \cos z
\sin r \sin \phi  ,
\nonumber
\\
u_{3} = \sin z  , \  u_{0} = \cos z \cos r  , \;\;
u_{1}^{2} +  u_{2}^{2} + u_{3}^{2} + u_{0}^{2}  = 1  .
\label{5.10a}
\end{eqnarray}
\noindent
In particular, we have
\begin{eqnarray}
\hspace{-3mm}r =0 , \ u_{1}=0, \  u_{2}=0,\ u_{3} = \sin z
 , \   u_{0} =  +\cos z   ,
\nonumber
\\
\hspace{-3mm}r =\pi   , \ u_{1}=0, \ u_{2}=0,\  u_{3} = \sin
z  , \   u_{0} =  -\cos z   ,\nonumber
\\
z \in [ - \pi/2 , + \pi
/2 ]
\label{5.10b}
\end{eqnarray}
\noindent which means that the full closed curve $ u_{0}^{2} +
u_{3}^{2} =1 $ in $S_{3}$ is parameterized by two pieces:
\begin{eqnarray}
\{  (r=0, \phi - \mbox{mute}, z) +  (r= \pi , \phi -
\mbox{mute}, z)  \}  .
\nonumber
\end{eqnarray}
\noindent Correspondingly, when  $m \neq 0 $,   the function  $R(r)$ must vanish in   $r=0, \pi$:
\begin{eqnarray}
m \neq 0\; , \qquad R(0) = 0\; , \qquad R(\pi) = 0 \; .
\label{5.10c}
\end{eqnarray}

First, let us examine the case $m=0$, then
\begin{eqnarray}
 a=0 , \ b = + B ,
\nonumber\\
 \lambda   + {1 \over 4} = 2B(n+1/2) + (n+1/2)^{2}  ,
\nonumber
\\
R =( \cos { r \over 2} )^{+2B}
{_2}F_{1}(-n , 2B + n + 1;  1 ;  - \sin^{2}{r \over 2} )  ,
\nonumber\\
  R _{r \rightarrow  0 } = 1 , \
R _{r \rightarrow  \pi } = 0  .
\label{5.11}
\end{eqnarray}
\noindent Because  $m=0$ and the wave function does not depend on  $\phi$,  this solution
is acceptable by continuity reasons.
Now, let us  examine the case $m>0$, then
\begin{eqnarray}
a =  + {m \over 2}, \ b = +
{m +2 B \over 2}  ,
\nonumber
\\
\lambda   + {1 \over 4} = 2B(n+m +1/2) + (n +m +1/2)^{2}  ,
\nonumber
\\
R =
  (\sin { r \over 2} )^{+m}
(\cos { r \over 2} )^{m+2B} \nonumber\\
\times  {_2}F_{1}(-n , 2B + 2m + n + 1;  m + 1 ;   - \sin^{2}{r
\over 2} ) \; ,
\nonumber
\\
  R _{r \rightarrow  0 } = 0  , \  R _{r \rightarrow  \pi } = 0  ;\quad
\label{5.12}
\end{eqnarray}
\noindent these are the correct single-valued solutions.
Let us consider the variant  $m < -  2B$, then
\begin{eqnarray}
a = -{m \over 2}   , \
b = -{m+2B \over 2}   > 0   ,
\nonumber
\\
\lambda   + {1 \over 4} = - 2B(n -m +1/2) + (n -m  +1/2 )^{2}
\nonumber
\\
= (n -m  +1/2 ) ((n -m  +1/2  -2B) > 0 \; ,
\nonumber
\\
R = ( \sin {r \over 2})^{ -m}   ( {\cos} {r \over
2})^{ -(m+2B)}  \nonumber
\\
\hspace{-2mm}\times {_2}F_{1}(-n  , -2m -2B +1 + n; -m  +1 ;  - {\sin}^{2}{r
\over 2} )  ,
\nonumber
\\
  R _{r \rightarrow  0 } = 0  , \
R _{r \rightarrow  \pi } = 0  .
\label{5.13}
\end{eqnarray}

Let us examine the case $-2B < m  < 0$, then
\begin{eqnarray}
a =  -{m \over 2}>0  ,
\ b = {m + 2B \over 2}  > 0  ,
\nonumber
\\
\lambda   + {1 \over 4} = 2B(n+1/2) + (n+1/2)^{2}  ,
\nonumber
\\
R =
 ({\sin} { r \over 2} )^{-m}
 ({\cos} { r \over 2} )^{m+2B}
\nonumber
\\
\times   {_2}F_{1}(-n , 2B + n + 1;  -m +  1 ;   - {\sin}^{2}{r
\over 2} )  ,
\nonumber
\\
 R _{r \rightarrow  0 } = 0 , \
R _{r \rightarrow  \pi } = 0  .
\label{5.14}
\end{eqnarray}

And finally, the last variant is $ m = -  2B$, then
\begin{eqnarray}
a = + B   , \ b =  0
 ,
\nonumber
\\
 \lambda   + {1 \over 4} = 2B(n+1/2) + (n+1/2)^{2} ,
\nonumber
\\
R = (- {\sin} {r \over 2})^{ +2B} \nonumber
\\
\times {_2}F_{1}(-n , 2B + n +1 ; -B
+1  ;   - {\sin}^{2}{r \over 2} )  ,
  \nonumber
  \\
  R _{r \rightarrow + 0 } = 0  , \ R _{r \rightarrow + \pi } = 1  .
\label{5.15}
\end{eqnarray}
\noindent We faced here with the case of discontinues  of wave functions in  $S_{3}$,
because the wave function preserves dependence on $\phi$ at $ r \rightarrow \pi $.

Collecting results together, we have
\begin{eqnarray}
m > 0  ,  \qquad
 \lambda +{1\over 4}=
(n+1/2 +m)(n+1/2+m+2B) ;
\nonumber
\\
m < -2B  ,  \qquad
 \lambda +{1\over 4}= (n+1/2 -m)(n+1/2-m-2B) ;
\nonumber
\\
-2B < m \leq 0  ,  \qquad]
 \lambda
+{1\over 4} = (n+1/2 )(n+1/2-2B)  .
\label{5.16}
\end{eqnarray}
\noindent
In usual unites, these formulas look
\begin{eqnarray}
\nonumber
m > 0  \,   \qquad \rho^{2}\lambda_{0} +{1 \over 4}
 = + 2{eB  \over \hbar c}\rho^{2}
(n+m+1/2) +(n+m+1/2)^{2} \; ;
\nonumber
\end{eqnarray}

\begin{eqnarray}
\nonumber
m < -2 {eB  \over \hbar c}  \rho^{2} \, \qquad  \rho^{2}\lambda_{0}+{1 \over 4}    =
-2 {eB  \over \hbar c}\rho^{2} (n-m+1/2) +(n-m+1/2)^{2}  ;
\nonumber
\end{eqnarray}
\begin{eqnarray}
 -2{eB  \over \hbar c}  \rho^{2} < m \leq
0 , \qquad
 \rho^{2}\lambda_{0} +{1 \over 4}   = 2 {eB  \over \hbar c} \rho^{2}(n+1/2) +(n+1/2)^{2} \; .
\end{eqnarray}
When  $\rho \rightarrow 0,
$ the case  $m < - \infty $ turns to be vanishing, and two other cases give
\begin{eqnarray}
 m < 0\; ,   \qquad
  \lambda_{0}  = 2 {eB  \over \hbar c} (n+1/2) ,
\nonumber
\end{eqnarray}
\begin{eqnarray}
m \geq 0 \; ,  \qquad \lambda_{0} = + 2{eB
\over \hbar c} (n+m+1/2)   .
\label{5.18}
\end{eqnarray}
\noindent Thus, we arrive at the known result in the flat space model:
\begin{eqnarray}
\lim_{\rho \rightarrow \infty} \lambda_{0} = {2M \over
\hbar^{2}} \left( E -{P^{2} \over 2M}\right)  ,\qquad
 E -{P^{2} \over 2M} =
 {eB \hbar \over M c} ( {m + \mid m \mid \over 2 } + n + 1/2)  .
\label{5.18}
\end{eqnarray}

\vspace{0.3cm}

\section{ Solutions of the equation for  $Z(z)$  in the space $H_{3}$}

Now, let us consider the differential equation for
 $Z(z)$  in space $S_{3}$:
\begin{eqnarray}
{d^{2} Z\over dz^{2}}  -  2\;  { \sin\; z \over \cos\;
z } \; {d Z \over dz}    + 2\epsilon \; Z - {\lambda  \over
\cos^{2} z} \; Z = 0 \; .
\label{6.1}
\end{eqnarray}
\noindent Changing the variable $\sin z = y ,$ we get
\begin{eqnarray}
(1-y^{2})  {d^{2} Z \over dy^{2}} - 3y {d  Z \over dy}
+ \left(
2\epsilon -{ \lambda \over 1- y^{2}} \right) Z = 0 .\qquad
\label{6.3}
\end{eqnarray}
\noindent In the variable $y =  (2x-1) $
it takes the form
\begin{eqnarray}
x(1-x) {d^{2} Z \over d x^{2}} - {3 \over 2}  (2 x-1)    {d
Z\over d x}
 + \left( 2\epsilon -{ \lambda \over  4x(1-x)} \right)Z = 0
. \label{6.5}
\end{eqnarray}
\noindent
With the use of the substitution
$
Z = x^{a} (1-x)^{b} F \; ,
$
 eq.  (\ref{6.5}) leads  to
\begin{eqnarray}
+ x (1-x) F'' +
 [  2a  (1-x)   - 2b x      -3x  +  {3 \over 2} ]   F'
+ [  a (a-1)+  {3 a/
2}   - {\lambda /
 4}    ]
{1\over x} F
\nonumber
\\
+ [   b(b-1) +3b  -  {3 b/
2} - {\lambda
/
 4} ] {1 \over 1-x}  F
+ [ -a(a-1)  - a b   - ab     - b(b-1)    - 3 a    - 3b
+ 2\epsilon ]   F    = 0  .
\nonumber
\\
\label{6.7a}
\end{eqnarray}
\noindent Imposing the following restrictions
\begin{eqnarray}
a^{2} + {a\over 2} - {\lambda \over 4}= 0\; , \qquad a = {-1 \pm
\sqrt{1+4\lambda} \over 4}\; ,
\nonumber
\\
 b^{2} + {b\over 2} - {\lambda \over 4}= 0\; ,\qquad
b = {-1 \pm \sqrt{1+4\lambda} \over 4}\;
\label{6.7b}
\end{eqnarray}
\noindent we reduce eq.   (\ref{6.7a})  to
\begin{eqnarray}
x (1-x)  F'' +  [   (2a  +3/2) -x (2a +2b +3)  ]   F'
-
 [ (a+b+1)^{2}     - 2 \epsilon  -1 ]  F    = 0 ,
\label{6.8a}
\end{eqnarray}
\noindent which is of hypergeometric type
\begin{eqnarray}
x(1-x) F + [ C - (A + B +1) x ] F' - AB  \; F = 0 \; ,
\nonumber
\end{eqnarray}
\noindent with parameters given by (to  obtain the single-valued wave functions we must take
$a$ and $b$ positive)
\begin{eqnarray}
Z = x^{a} (1-x)^{b}\; {_2}F_{1}(A, B; C; x)  ,\qquad
  a = {-1 +  \sqrt{1+4\lambda} \over 4}, \qquad
 b = {-1  + \sqrt{1+4\lambda} \over 4} ;
\nonumber
\\
 C =  1 +  \sqrt{\lambda +1/4} , \qquad
 A =  {1 \over 2}  +  \sqrt{\lambda+{1\over 4}}   +  \sqrt{2\epsilon+1} ,
 \qquad
  B =  {1 \over 2}   +   \sqrt{\lambda+{1\over 4}}   -
\sqrt{2\epsilon+1}
 .
\nonumber
\\
\label{6.11}
\end{eqnarray}
\noindent To obtain polynomials we must impose the following conditions:
\begin{eqnarray}
\lambda + {1 \over 4} > 0 , \quad 2\epsilon +1 > 0 ,
\quad B =  {1 \over 2}   \nonumber\\
+   \sqrt{\lambda+{1\over 4}}   -
\sqrt{2\epsilon+1}   = -N  , \ N = 0, 1, 2, \ldots
\nonumber
\end{eqnarray}
\noindent from whence  it follows the  quantization rule
\begin{eqnarray}
  2\epsilon+1
  = \left({1 \over 2}   + N +    \sqrt{\lambda+{1\over 4}}  \right)^{2} \; .
\label{6.12}
\end{eqnarray}
\noindent In particular, this formula gives
\begin{eqnarray}
  (2\epsilon+1) - \left(\lambda+{1\over 4}\right)
  = \left({1 \over 2}   + N \right)^{2}
+   2  \left({1 \over 2}   + N \right) \sqrt{\lambda+{1\over 4}}
   ,
\label{6.13}
\end{eqnarray}
\noindent
with  the notion
$$
\left({1 \over 2}   + N \right)^{2} +   2  ({1 \over 2}   + N )
\sqrt{\lambda+{1\over 4}} = \nu^{2} > 0 
 $$
\noindent it reads
\begin{eqnarray}
  (2\epsilon+1) =   \nu^{2} + (\lambda+{1\over 4})
 \label{6.14b}
\end{eqnarray}
\noindent which generalizes the known relation in flat space model:
$  2\epsilon  =   k^{2} + \lambda    $.
The identity   (\ref{6.14b}) permits us to express all parameters through
$\epsilon, \nu^{2}$:
\begin{eqnarray}
 C =  1 +  \sqrt{(2\epsilon +1 )- \nu^{2}}, \nonumber\\
 A =  {1 \over 2}  +   \sqrt{(2\epsilon +1 )- \nu^{2}}   +  \sqrt{2\epsilon+1}
   ,
\nonumber
\\
B =  {1 \over 2}   +   \sqrt{(2\epsilon +1 )- \nu^{2}}   -
\sqrt{2\epsilon+1}    .
\label{6.15}
\end{eqnarray}
\noindent However, it should be noted that  such a form does nod provide us with
any  advantage because quantization for  the quantity  $(2\epsilon +1 )- \nu^{2}$ is given by the differential
equation for radial function
$R(r)$.
It is easily to obtain an operator
with eigenvalues $\nu^{2} +3/4$:
\begin{eqnarray}
\hat{\nu}^{2} -{3\over 4} = 2 \hat{H} (r,\phi,z)   - \hat{\lambda}
(r),
\nonumber
\\
\hat{\lambda} (r, \phi) =
 -
{1 \over  {\sin} r}  {\partial \over  \partial r }
{\sin}\;  r \;
  {\partial \over  \partial r } \;  +
{1 \over {\sin}^{2} r   }   \left[\; -i {\partial \over
\partial \phi  } +
 B (1 -{\cos} r )  \right] ^{2}  ,
 \label{6.16}
\end{eqnarray}
\noindent so that the operator is given by
\begin{eqnarray}
&\hat{\nu}^{2}(r,\phi,z)   -{3\over 4} = \left({  1\over {\cos}^{2}z
} - 1\right) \hat{\lambda} (r,\phi) - {1 \over {\cos}^{2} z }\nonumber\\[2mm]
&\times {\partial \over  \partial z}  {\cos}^{2} z { \partial \over
\partial z}
   .
\label{6.17}
\end{eqnarray}
\noindent It commutes with Hamiltonian.

\section{  Presence of an electric field, the case of   space $S_{3}$}

Let us introduce into consideration an additional external electric field along the $z$ axis

\begin{eqnarray}
A_{0} = E_{0} \; {\sin z \over  \cos^{2} z}
\label{7.1}
\end{eqnarray}

\noindent in the limit of flat space it becomes the uniform constant field
 $E$ along  $z$. As we will see below, this additional field does not destroy
hypergeometric structure of the solution $Z(z)$.
To take into account the  presence of this field, it is  sufficient to
make one  formal change in the previous calculations
(in dimensionless notation)
\begin{eqnarray}
 \epsilon
 \Longrightarrow  \epsilon + eE_{0} {  M \rho^{2} \over
\hbar^{2} } \rho  {  \sin z \over  \cos^{2} z} 
=   \epsilon
+ \mu  { \sin z \over  \cos^{2} z}  .
\label{7.3}
\end{eqnarray}
\noindent Also, one should use the identity
\begin{eqnarray}
(  2\epsilon  + 2 \mu  { \sin z \over  \cos^{2} z} ) = 2\epsilon
+ 2 \mu { y \over 1-y^{2}} =
 2\epsilon
 + 2 \mu { 2x-1  \over 4x (1-x)} = 2\epsilon+ {2\mu
\over 4} ( - {1 \over x} + {1 \over 1-x})  .
\nonumber
\label{7.4}
\end{eqnarray}
\noindent Correspondingly, instead of  eq.~(\ref{6.7a}), we have
\begin{eqnarray}
x (1-x) F'' +
 [  2a  (1-x)   - 2b x      -3x  +  {3 /
 } ]   F'
\nonumber
\\
+ \left[  a (a-1)+  {3 a\over 2}   - {\lambda \over 4}  - {2\mu \over
4}  \right] {1\over x} F + \left[   b(b-1) +3b  -  {3 b\over 2} -
{\lambda \over 4} + {2\mu \over 4}   \right] {1 \over 1-x}  F +
\nonumber
\\
+ [ -a(a-1)  - a b   - ab     - b(b-1)    - 3 a    - 3b
+ 2\epsilon ]   F    = 0  .
\label{7.5}
\end{eqnarray}
\noindent By imposing two conditions
\begin{eqnarray}
a^{2} + {a\over 2} - {\lambda + 2\mu \over 4}= 0 , \
 b^{2} + {b\over 2} - {\lambda -2\mu  \over 4}= 0
\label{7.6}
\end{eqnarray}
\noindent eq.   (\ref{7.5})  reduces to
\begin{eqnarray}
 x (1-x)  F'' +
 [   (2a  +3/2) -x (2a +2b +3)  ]   F'
 -
  [ (a+b+1)^{2}     - 2 \epsilon  -1 ]   F    = 0
\label{7.7a}
\end{eqnarray}
\noindent which is of hypergeometric type. Thus the solutions are given by
\begin{eqnarray}
Z = x^{a} (1-x)^{b} {_2}F_{1}(A, B; C; x)   ,
\qquad
 x = {1 +  {\sin} z \over 2},  \qquad  (1- x) = {1 -   {\sin} z \over 2} ,
\nonumber
\\
 a = {-1 +  \sqrt{1+4(\lambda+2\mu)} \over 4} ,\qquad
 b = {-1  + \sqrt{1+4(\lambda - 2\mu) } \over 4}  ,
\nonumber
\\
  A =  {1 \over 2}  +  {1 \over 2} \sqrt{(\lambda +2\mu) +{1\over 4}}  +
  {1 \over 2}  \sqrt{(\lambda-2\mu) +{1\over 4}}    +  \sqrt{2\epsilon+1}  ,
\nonumber
\\
B =  {1 \over 2}   +  {1 \over 2} \sqrt{(\lambda +2\mu) +{1\over
4}} + {1 \over 2} \sqrt{(\lambda-2\mu) +{1\over 4}} -
\sqrt{2\epsilon+1},
\end{eqnarray}
\begin{eqnarray}
 C =  1 +  \sqrt{(\lambda -2\mu)  +1/4}
 .
\nonumber
\end{eqnarray}

To obtain polynomials, we must impose the following restriction
\begin{eqnarray}
(\lambda \pm 2\mu)  + {1 \over 4} > 0, \ 2\epsilon +1 >
0 ,
\nonumber
\\
  B =  {1 \over 2} +   \sqrt{(\lambda +2\mu) +{1\over 4}} +
\sqrt{(\lambda-2\mu) +{1\over 4}}
 - \sqrt{2\epsilon+1}
  = -N  ,\ N = 0, 1, 2, \ldots
\nonumber
\end{eqnarray}

\noindent from whence it follows the  quantization rule

\begin{eqnarray}
  2\epsilon+1
  = \Big( {1 \over 2}   + N +   {1 \over 2} \sqrt{(\lambda +2\mu) +{1\over 4}}
+
{1 \over 2} \sqrt{(\lambda-2\mu) +{1\over 4}}    \Big)^{2}
\label{7.10}
\end{eqnarray}
\noindent where  the quantization of $\lambda$ is  given by  (\ref{5.16})).
Thus, the energy levels are determined by two quantum  numbers,
$(n,N)$,  and  depends upon three parameters, curvature radius $\rho$,
and magnitudes of fields:  $ B$  and  $E$.

\label{last}
\end{document}